\documentclass[%
 aip,
preprint,%
]{revtex4-1}

\usepackage{natbib}
\usepackage{lineno}
\usepackage{floatflt}

\usepackage[normalem]{ulem}

\usepackage[dvips]{graphicx}% Include figure files
\usepackage{dcolumn}% Align table columns on decimal point
\usepackage{bm}% bold math
\usepackage[all]{xy}
\usepackage{amssymb}
\usepackage{amsmath}

\usepackage{color}

\begin{document}

\begin{abstract} 
We chemically characterize the symmetries underlying the exact
solutions of a stochastic negatively self-regulating gene. The
breaking of symmetry at low molecular number causes three effects.
Two branches of the solution exist, having high and low switching
rates, such that the low switching rate branch approaches
deterministic behavior and the high switching rate branch exhibits
sub-Fano behavior. Average protein number differs from the
deterministically expected value. Bimodal probability distributions
appear as the protein number becomes a readout of the ON/OFF state of
the gene.
\end{abstract}

\preprint{AR-4}
% Physics and Astronomy Classification Scheme.
% Use showkeys class option if keyword

\title{Physical implications of $\mathfrak{so}(2,1)$ symmetry in exact solutions for a self-repressing gene}

\author{Alexandre F. Ramos}

\affiliation{Escola de Artes, Ci\^encias e Humanidades, Departamento de Radiologia e Oncologia -- Faculdade de Medicina, Universidade de S\~ao Paulo -- Instituto do C\^ancer do Estado de S\~ao Paulo -- Av. Arlindo B\'ettio, 1000 CEP 03828-000, S\~ao Paulo, SP, Brazil}

\email{alex.ramos@usp.br}

\author{John Reinitz}

\affiliation{Departments of Statistics, Ecology \& Evolution, 
Molecular Genetics \& Cell Biology,  Jones Laboratory, University of Chicago, 
5747 South Ellis Ave, Chicago, IL 60637, USA}

\maketitle

\date{today}
  
  {Symmetries, described by Lie algebras, have been central tools for new
    discoveries in quantum mechanics and quantum field theory
    \cite{Anderson1958,Gellmann1964,Higgs1964}. Applications of these techniques
    to problems in statistical physics have been limited
    \cite{Cannavacciuolo2016,Tamm2016}. Over the last two decades, statistical
    physics has been widely applied to biological systems \cite{Arkin1998}. Such
    studies frequently make use of the chemical master equation (CME), typically
    solved by Gillespie's direct simulation method \citep{Gillespie1977}.  This
    method requires reconstructing probability distributions experimentally from
    repeated computational runs, a procedure that can overlook important
    features of the distributions. In a case where exact solutions were obtained
    to the CME for a self-repressing gene \cite{Hornos2005} in terms of
    generating functions with a symmetry described by a $\mathfrak{so}(2,1)$ Lie
    algebra \cite{Ramos2007}, physical insight into the general behavior of the
    system was limited by the dimensionality of the parameter space and the lack
    of a physical interpretation of the symmetry. In this paper, we show that
    this symmetry has three important physical effects. First, an invariant
    quantity under the group's action is a quadratic function of a certain ratio
    of protein removal rates. The two roots of that quadratic induce two
    branches of the solution. One branch approaches a deterministic regime as
    molecular number increases, while the other represents a novel class of
    stochastic behavior. Second, symmetry in the form of numerical equivalence
    between the average number of molecules in the system in the deterministic
    and stochastic regimes is lost when molecular number becomes sufficiently
    small. Finally, actions of the group that leave the system unchanged in the
    deterministic limit have two completely distinct effects in the stochastic
    regime depending on whether the system can be defined by a Langevin
    approximation or not. These physical manifestations of the underlying
    symmetry also provide a systematic characterization of the model's behavior
    in the entire parameter space of the exact solutions. This mathematical
    characterization has previously been reported by use of the Poisson
    representation \cite{Gardiner1977,Biswas2014}. The analysis of symmetries
    reveals new phenomenology in a fundamental physical system for investigating
    gene networks. The use of group theoretical techniques on generating
    functions represents a new class of applications of this technique for
    related problems involving the CME \cite{Innocentini2007,Lepzelter2008} and
    other applications \cite{Hornos1993,Bashford2004}. It has provided us with a
    way to identify the building blocks necessary to understand the workings of
    randomness and invariance in biological systems.
        
We conceive a deterministic model for a negative self-regulating gene
as an ensemble of genes (operators, in the case of a prokaryote) at
concentration $[O_T]$. The operators may be in the ON or OFF state if
they are, respectively, unbound or bound to the regulatory
protein. The concentration of ON (OFF) operators is indicated by $[O]$
($[OP]$), with $[O_T]=[O]+[OP]$, and the protein concentration is
given by $[P]$. The macroscopic reaction scheme is given by
\begin{equation}\label{eq:chmrctn}
  \begin{tabular}{lccc}
   {\rm Protein synthesis}: & O   & $\stackrel{\hat{k}}{\rightharpoonup}$    & P+O \\
   {\rm Protein decay}:     & P   & $\stackrel{\hat{\rho}}{\rightharpoonup}$ & $\varnothing$ \\
   {\rm Switching OFF}:     & P+O & $\stackrel{\hat{h}}{\rightharpoonup}$    & OP \\
   {\rm Switching ON}:      & OP  & $\stackrel{\hat{f}}{\rightharpoonup}$    & P+O.
  \end{tabular}
\end{equation}

All macroscopic rate constants are written with
hats.  {$\hat{k},$ $\hat{\rho},$ and $\hat{f}$ each have units
of minute$^{-1}$, while $\hat{h}$ has units of liter/minute.}   
(\ref{eq:chmrctn}) implies that
\begin{eqnarray}
  \frac{d[P]}{dt} &=&
  \hat{k}[O] + \hat{f} [OP] - (\hat{\rho} + \hat{h}[O])[P], \nonumber \\
  \frac{d[O]}{dt} &=&
  \hat{f}[OP] - \hat{h}[O][P], \nonumber \\
  \frac{d[OP]}{dt} &=&
  -\hat{f}[OP] + \hat{h}[O][P].\nonumber
\end{eqnarray}
 {We denote the steady state concentrations of $[O],$
$[P],$ and $[OP]$ by $\overline{O},$ $\overline{P},$ and $\overline{OP}.$ Then
\begin{eqnarray} 
  \frac{\overline{O}}{ [O_T]} &=&
  \frac{\hat{f}}{\hat{f}+\hat{h} \overline{P}} =
  \frac{1}{1+K \overline{P}}, \ \ \
  \frac{\overline{OP}}{[O_T]} = 1-\frac{\overline{O}}{[O_T]}, \label{eq:stdconc1} \\
  {\overline{P}} &=& N {\overline{O}} = \frac{\sqrt{1+4KN[O_T]}-1}{2K}, 
\label{eq:stdconc2}
\end{eqnarray}
w}here $K$, the equilibrium affinity, is given by
$K=\frac{\hat{h}}{\hat{f}}$  {liters}, and
$N=\frac{\hat{k}}{\hat{\rho}}$. Eq. (\ref{eq:stdconc1}) indicates that
the  {rate at which} operators  { move
  from} OFF to ON and ON to OFF is, respectively, given by $\hat{f}$
and $\hat{h} {\overline{P}}$. Thus, the total
 {rate of operator switching} in both directions is
$\hat{f}+\hat{h} {\overline{P}}$.  Note that the
expected  {concentration} of proteins is given by the
product of the ratio between the protein synthesis and degradation
rates and the concentration of ON operators. Hence, at the limit of
small affinity of the repressor for the operator ($K \rightarrow 0$),
 {$\overline{P}/[O_T] = N$, the expected number of
  proteins in the absence of regulation}.

A stochastic model for the negative self-regulating gene has been
proposed in terms of two random variables, the protein number, denoted
by $n$, and the operator state, which can be ON or OFF
\cite{Hornos2005}.  The  {steady state} probability of
finding $n$ proteins  {and} the operator
ON or OFF  {is denoted by $\alpha_n$ or $\beta_n$,}
respectively. In the stochastic model, we replace the reaction rate
constants of Eq. (\ref{eq:chmrctn}) by propensities represented by the
unhatted symbols  {$k=\hat{k},$ $\rho = \hat{\rho},$ $f
= \hat{f},$ and $h = V \hat{h},$ where $V$ is the system volume 
\cite{Gillespie1977}.}
Note that now we may consider
a single gene instead of an ensemble and the proportion of ON
operators of the deterministic model becomes the marginal probability
of finding the operator ON, $P_\alpha = \sum_{n=0}^\infty
\alpha_n$. 
 {T}he marginal probability of
finding $n$ proteins in the cytoplasm independently of the operator
state being ON or OFF is given by $\phi_n=\alpha_n+\beta_n$ and is
computed in terms of the KummerM functions, so that
\begin{equation}     \nonumber
  \phi_n =
  \frac{(Nz_0)^n}{ c \, n!}
  \frac{(a)_n}{(b)_n}
  \, {\rm M} (a+n,b+n,-Nz_0^2),
\end{equation}
where  {$(x)_n$ denotes the Pochhammer symbol defined by $(x)_n=x(x+1)\dots(x+n-1)$ and $(x)_0=1$, and}
\begin{eqnarray}
  c   = {\rm M} (a,b,Nz_0(1-z_0)), \ \       
  z_0 = \frac{\rho}{\rho+h}, \label{eq:Nz0}\\
  N   = \frac{k}{\rho}, \ \                      \nonumber
  a   = \frac{f}{\rho}, \ \   \nonumber
  b   = \frac{f}{\rho+h}+\frac{h\,k}{(\rho+h)^2}. \nonumber %JR \label{eq:ab}
\end{eqnarray}
$N$ is the average number of proteins at the steady state regime if the operator
is fully ON.  $z_0$ gives the proportion of protein removal from cytoplasm by
first order decay. $a$ is the ratio of the OFF to ON transition rate to the
protein degradation rate. $a\gg 1$  {($a \ll 1$)} indicates a
regime where, on average, the OFF operator switches back to the ON state faster
 {(slower)} than the time required for protein degradation. $b$
gives the ratio of the operator switching to the protein removal rates. The
operator switching rate is the sum of the average OFF to ON switching rate $f$
and the ON to OFF rate defined in analogy with the deterministic case to be
$hk/(\rho+h)$.  For $b\gg1$ the operator switches multiple times between the ON
and OFF states during the average time for protein removal. In that case, the
probability distributions for protein number are unimodal. On average, for
$b \approx 1$ or smaller, the operator takes longer to switch from ON to OFF to ON
(or vice-versa) than the average protein removal time. In that case, and for $a
\approx hk/(\rho+h)$, the probability distributions for protein number are
bimodal because most of the proteins synthesized when the operator is ON decay
before the operator switches OFF.  { The average protein number
  of the distribution is given by
  \begin{equation} \nonumber
    \langle n \rangle = N \frac{a}{b}\frac{z_0}{c} {\rm M}(a+1,b+1,Nz_0(1-z_0)),  
  \end{equation}
which can be written as $\langle n \rangle = N P_\alpha$, with $P_\alpha$ the
stochastic equivalent of $\overline{O}$ in Eq. (\ref{eq:stdconc2}).  }

This stochastic model is a combination of two stochastic processes,
and hence approaches equilibrium at the two rates $\rho$ and
$b(\rho+h)$, the former related to the protein degradation and the
latter to operator switching \cite{Ramos2011}.  The smallest of those
two rates determine when the system reaches equilibrium.  The time
dependent solutions  {in terms of generating functions  has} 
the form
\begin{equation}
 {
\label{eq:Heun}
  \phi(z,t)
  \propto
      {\rm e}^{- j \rho t } {\mathcal H}_{1,j}(z) + {\rm
        e}^{-(\rho+h)(b+j)t} {\mathcal H}_{2,j}(z)},
\end{equation}
where $j$ is a non-negative integer and $ {\mathcal H}_{1,j} $ and $
{\mathcal H}_{2,j} $ are confluent Heun functions
\cite{Ramos2011,Ronveaux1995}.  {These solutions are 
  obtained applying the separability {\it ansatz} whose $z$ component obeys
  a second-order ODE having two regular poles, one
around $z=1$, giving ${\mathcal H}_{1,j}(z)$, and the other around
$z=z_0$, giving ${\mathcal H}_{2,j}(z)$. 
It is evident that the only steady state solutions in Eq. (\ref{eq:Heun}) have
vanishing time dependent exponents, implying the selection of $j=0$ and
${\mathcal H}_{1,0}(z)$.
${\mathcal H}_{1,0}(z)$ can then be written as a
a KummerM function, so that
\begin{equation} \label{eq:phiz}
  \phi(z)
  = c \, {\rm M}(a,b,Nz_0(z-z_0))
  = \phi_{b,a},
\end{equation}
is the generating function of the probabilities $\phi_n$.
}

These exact solutions of the steady state stochastic model indicated  the
existence of $\mathfrak{so}(2,1)$ symmetries \cite{Ramos2007}. 
The generating functions $\phi$ in Eq (\ref{eq:phiz}) 
span irreducible representations of $\mathfrak{so}(2,1)$, which
in the Cartan basis has its operators denoted by $H$ and $E_\pm$. The
Casimir operator is defined as $C=-H^2 + H + E_+E_-$ and the
commutation relations are
$$
[H,E_\pm] = \pm E_\pm, \ \ [E_+,E_-] = -H, \ \ [C,H]=[C,E_\pm]=0.
$$
The action of the algebraic operators on the generating functions
$\phi_{b,a}$ is:
\begin{eqnarray}
  C \, \phi_{b,a} = \left( \frac{1-b^2}{4} \right) \, \phi_{b,a}, \label{eq:csmr} \ \ \ \ \ \ \ \ \ \ \ \ \ \ \ \ \ \\ %\ \ \,
  H \, \phi_{b,a} = \left( \frac{2a+1-b}{2} \right) \, \phi_{b,a}, \label{eq:car1}  \ \ \ \ \ \ \ \ \ \ \ \ \\
  E_+ \, \phi_{b,a} = a \, \phi_{b,a+1}, \ \ \ \ %\\ %\ \ \ \ \ \
  E_- \, \phi_{b,a+1} = (b-a) \, \phi_{b,a}. 
  \label{eq:lad1}
\end{eqnarray}
The invariant of the algebra is determined by the eigenvalue of the
Casimir operator and Eq. (\ref{eq:csmr}) implies that $b$ is
constant. The Cartan operator's eigenvalue  {in}
Eq. (\ref{eq:car1}}) determines the OFF to ON switching rate in relation
to the protein degradation rate, while the ladder operators change the
value of $a$ by one.

\begin{figure}
  \centering
  \hfill
  \includegraphics[width=0.32\linewidth]{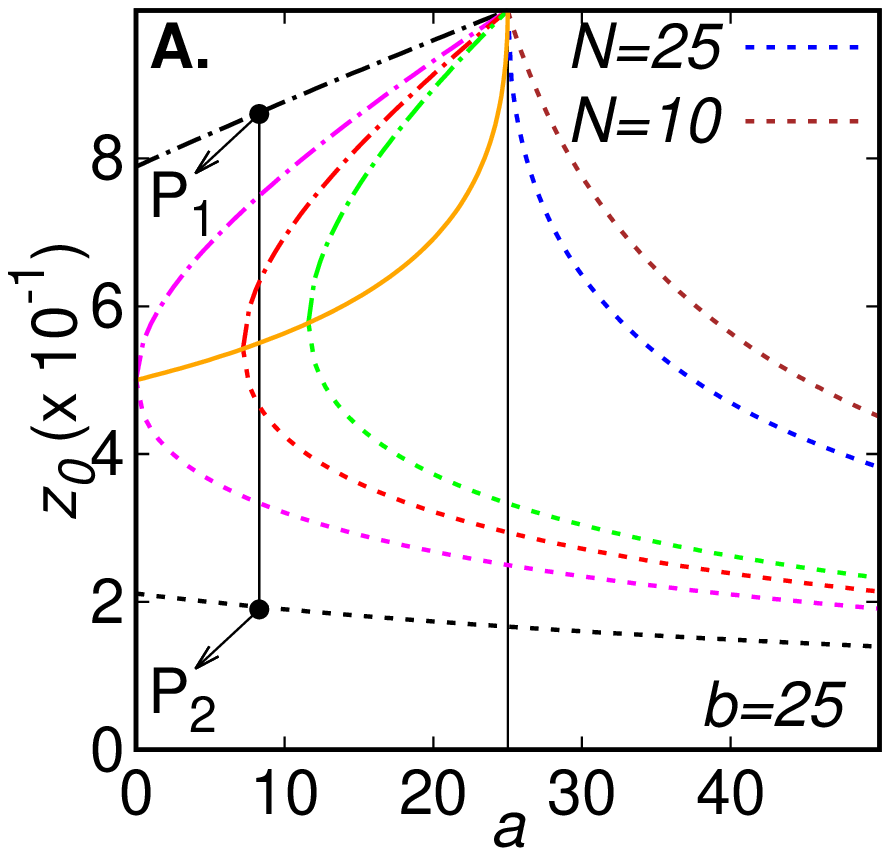}
  \hfill
  \includegraphics[width=0.32\linewidth]{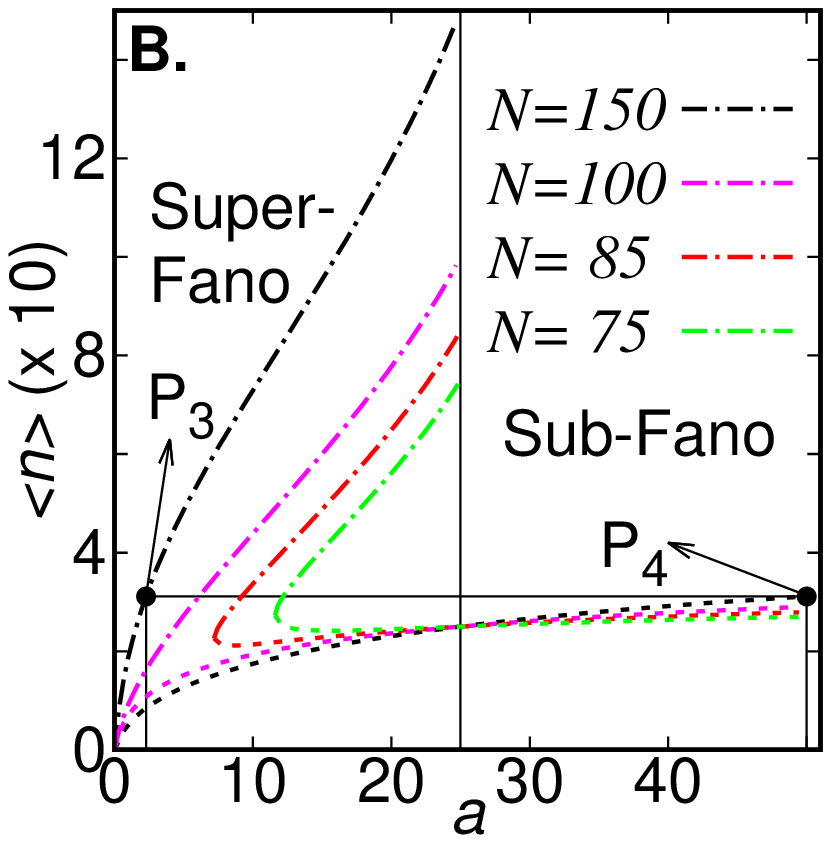}
  \hfill
  \includegraphics[width=0.32\linewidth]{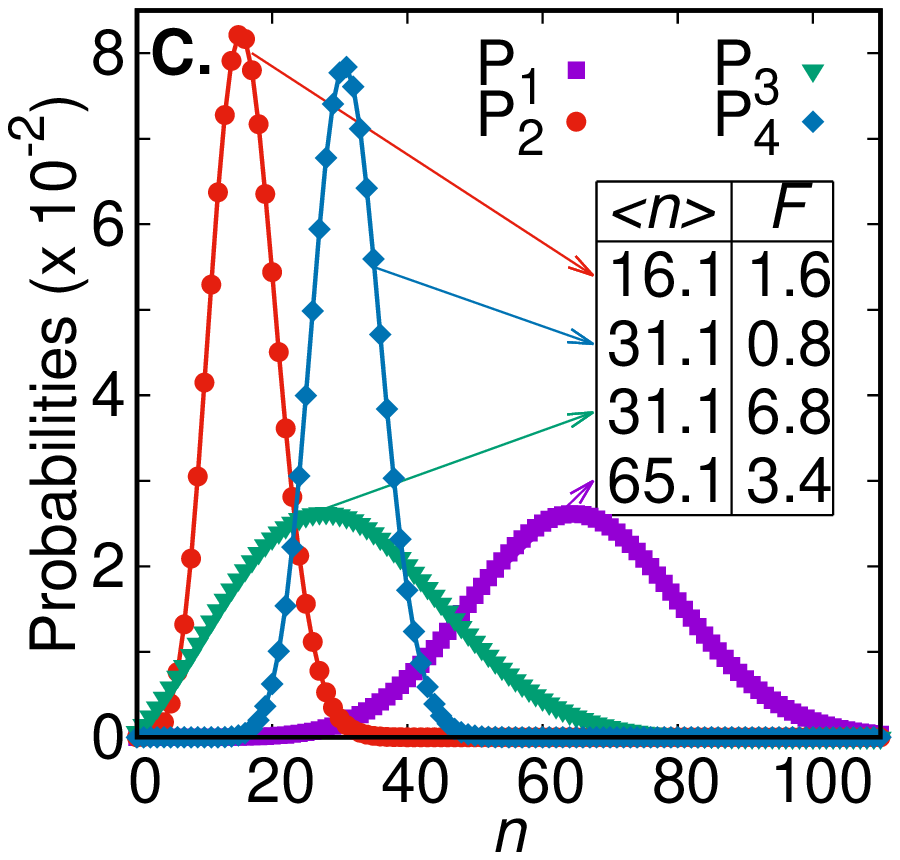}
  \hfill
  \caption{(A) and (B) show $z_0$ and $\langle n \rangle$, respectively, as
    functions of $a$ for fixed values of $N$ as indicated by the keys in A and
    B. Dashed-dot (dashed) lines correspond to $z_0^+$ ($z_0^-$). The vertical
    black line at $a=25$ separates the sub-Fano and super-Fano noise regimes of
    the steady state probability distribution. In graph A, the
     {tan solid} line indicates $z_0^+=z_0^-$, and $P_1$ and
    $P_2$ show the $(a,z_0)$ values for two distributions shown in (C). (B)
    shows the dependence of $\langle n \rangle$ on $a$. (C) shows
     {steady state} probability distributions with
    $(N,b)=(150,25)$.  $(a,z_0)$ for each distribution are P$_1=(8.3,0.86)$,
    P$_2=(8.3,0.19)$, P$_3=(2.2832,0.14)$, P$_4=(50,0.81)$ with $z_0$ calculated
    from Eq. (\ref{eq:z0}), where P$_1$, $P_3$ (or P$_2$, P$_4$) were calculated
    with $z_0^+$ (or $z_0^-$). The Fano factor of each probability distribution
    is indicated by $F$.}
  \label{fig:z0}
\end{figure}

We start building the biological interpretation of the symmetries of
the model by writing its invariant as $b=az_0 + Nz_0(1-z_0)$. A fixed $b$
leads to a 3D locus embedded in a 4D space. For fixed values of $N$
we obtain two possible values for $z_0$, given by
\begin{equation}\label{eq:z0}
  z_0^\pm = (1+a/N)/2 \left(1 \pm \sqrt{ 1 - 4bN(N+a)^{-2} } \right).
\end{equation}

FIG \ref{fig:z0}A shows the possible values of $z_0^\pm$ as functions
of $a$.  $a \ge b$ implies $z_0^+>1 $ which is biologically
meaningless and only $z_0^-$ has acceptable values
(Eq. \ref{eq:Nz0}). For a given $a\le b$ the
dynamical regime of the system is degenerate and two values of $z_0$
distinguish those regimes in terms of the ON to OFF operator state
transition. The first regime, ($z_0^-$), has strong self-repression
(high value of $h$) and low steady state protein number. The second
regime, ($z_0^+$), is characterized by a high steady state protein
number and weak self-repression (low value of $h$).

FIG \ref{fig:z0}B shows a further consequence of this degeneracy on
the average protein number. For sufficiently low $\langle n \rangle$,
one has two possible values of $a$ and $z_0^-$, both characterized by
the same value of $b$.  Those values indicate two regimes of operator
switching, with lower (or higher) values for the switching rates $f$
and $h$, that is, slow or fast switching. For the specific condition
when one regime has $a>b$ and the other has $a<b$ the noise on the
protein numbers is characterized, respectively, as sub-Fano and
super-Fano.

\begin{figure}
  \centering
  \includegraphics[width=0.45\linewidth]{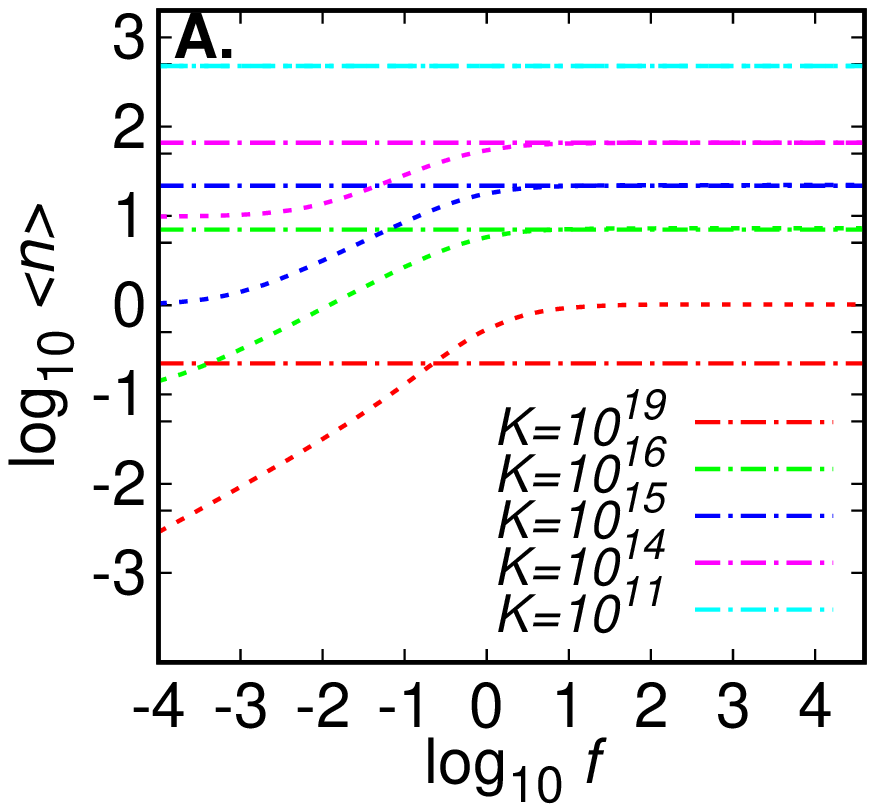}
  \hfill
  \includegraphics[width=0.45\linewidth]{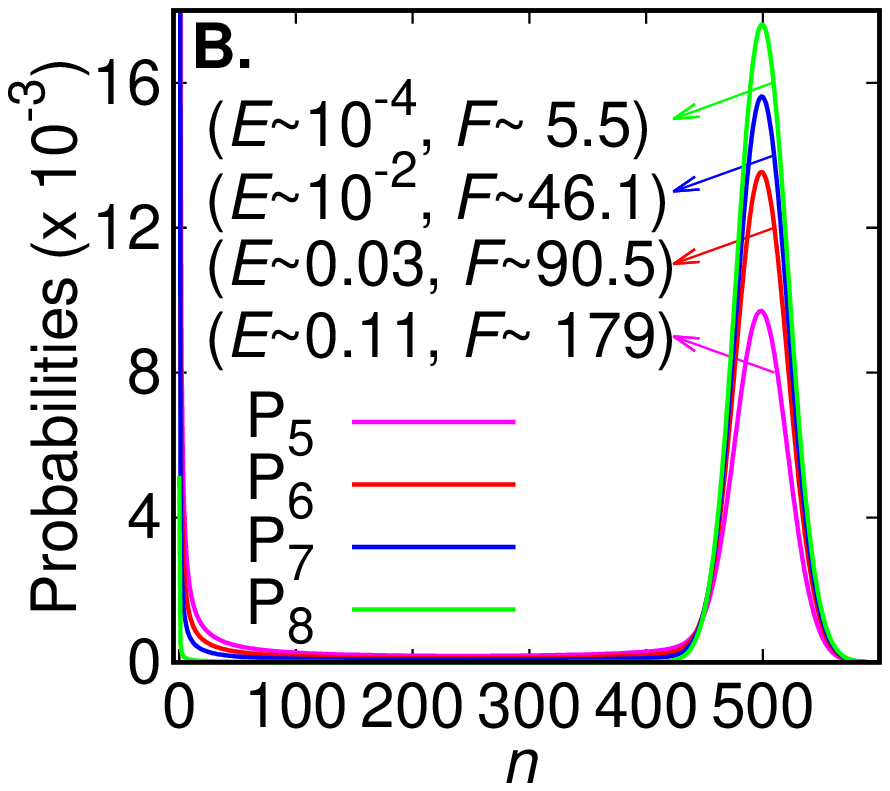}
  \caption{(A) shows a comparison of the expectation of the steady state protein
    number of the stochastic model  {(dashed lines)} and the
    protein number from the deterministic model  {(dashed-dot
      lines)} as a function of the parameter $f$.  { In the
      deterministic limit, $ \langle n \rangle = V \overline{P},$ and we set
      $V=10^{-15}$ liters, the volume of a bacterial cell. $K$, in units of
      liters, is shown in the key.}  Synthesis and degradation rates are
    $k=500$, $\rho=1$. (B) shows the distributions when $\langle n \rangle $ and
     {$\overline{P}$} are comparable.  The relative error $E$ is
    given by  { $ E = \frac{\|\langle n \rangle - \overline{P}
        \|}{ {\rm max}(\langle n \rangle, \overline{P})} $}.  The key shows the
    distributions by color and $E$ and the Fano factor $F$ are given for each
    distribution.  The parameters $(N,b)=(500, 0.1)$, and $(a,K)$ are
     {P$_5 = (0.06, 1.3 \times 10^{12})$, P$_6 = (0.08, 5\times
      10^{11})$, P$_7 = (0.09, 2\times 10^{11})$, P$_8 = (0.099, 2\times
      10^{10})$}. The probabilities of finding up to 400 proteins (or more than
    400) for curves P$_5$, P$_6$, P$_7$, and P$_8$, are approximately 0.42 (or
    0.58), 0.22 (or 0.78), 0.11 (or 0.89), 0.01 (or 0.99), respectively.}
  \label{fig:split}
\end{figure}

The stochastic model exhibits splitting between the deterministic and
stochastic solutions to the dynamics of the negative self-regulating
gene when the average protein numbers are low. FIG \ref{fig:split}A
shows a comparison between the steady state concentration of proteins
predicted by the deterministic model in Eq. (\ref{eq:stdconc2}) and
the average protein number as given by the stochastic model. For high
values of $\overline{P}$ there is a good agreement for the steady
state number of proteins predicted by both the stochastic and
deterministic approaches. As the steady state number of proteins
decreases, discrepancies between the two approaches start to appear as
$f\rightarrow 0$. For $h$ sufficiently high, the probability for the
gene being OFF increases and when $f$ becomes very small the
stochastic and deterministic solutions diverge.  The correspondence
principle breaks down when the molecular number is extremely small.
FIG \ref{fig:split}A shows that for  {$K=10^{19}$}, large
values of $f$ cause the protein number to approach $1$.  This is a
consequence of the fact that a protein bound to the operator does not
decay in the reaction scheme given in Eq (\ref{eq:chmrctn}), and we
have shown elsewhere that this case can give rise to Fano factors
arbitrarily close to zero \cite{Ramos2015}.

\begin{figure}
  \centering
  \hfill
  \includegraphics[width=0.32\linewidth]{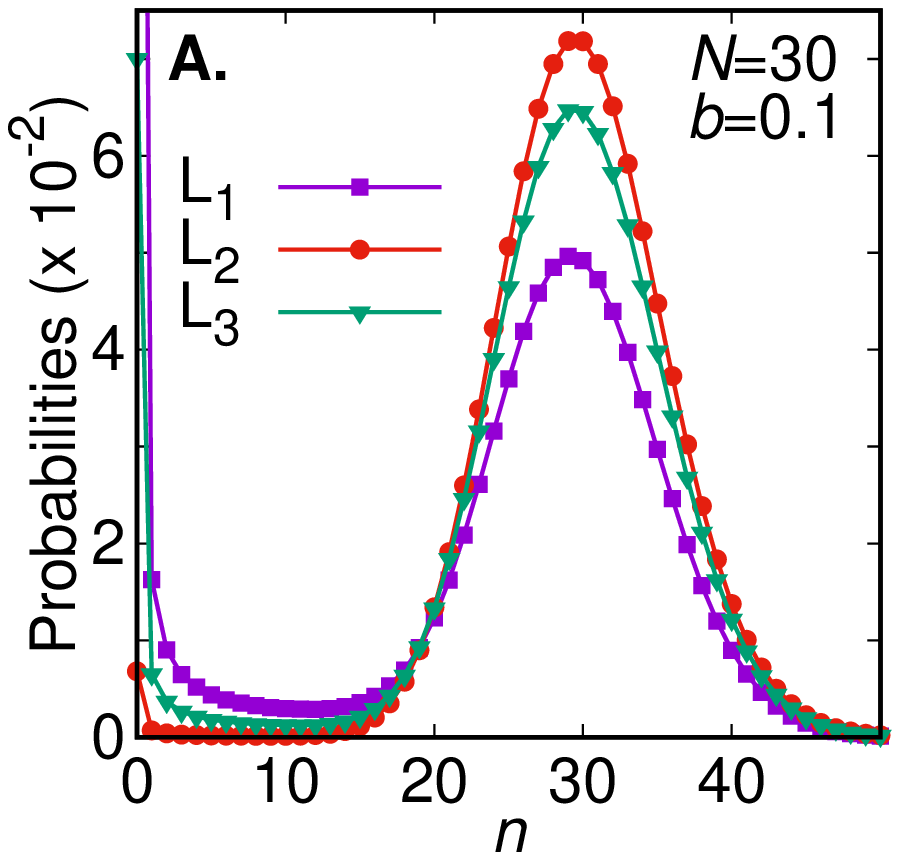}
  \hfill
  \includegraphics[width=0.32\linewidth]{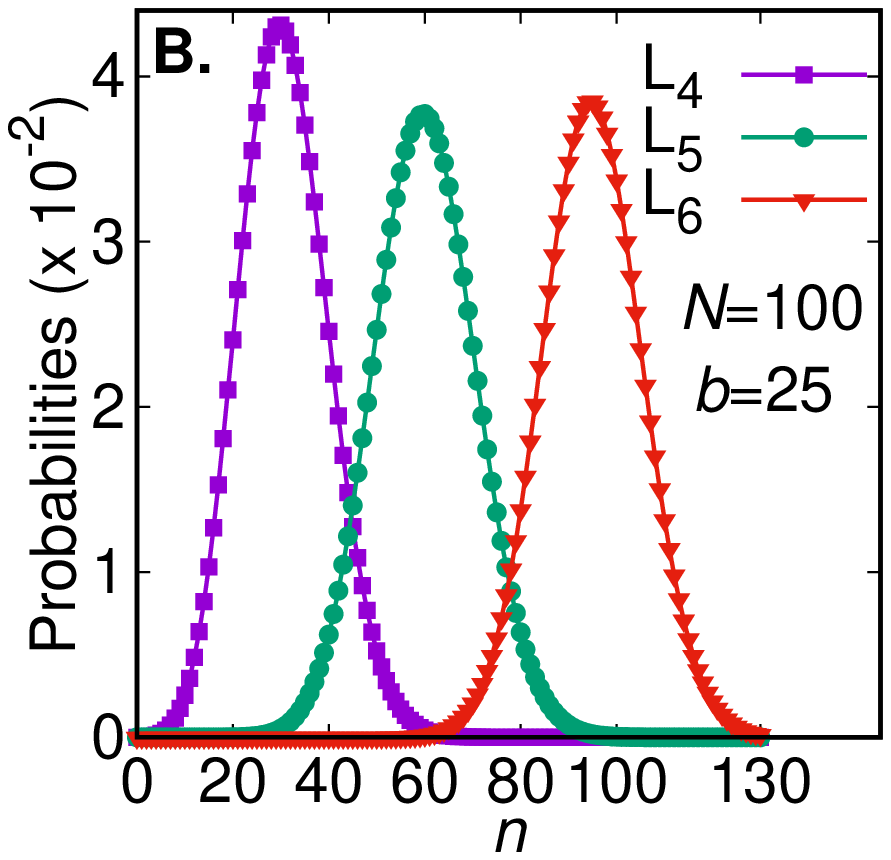}
  \hfill
  \includegraphics[width=0.32\linewidth]{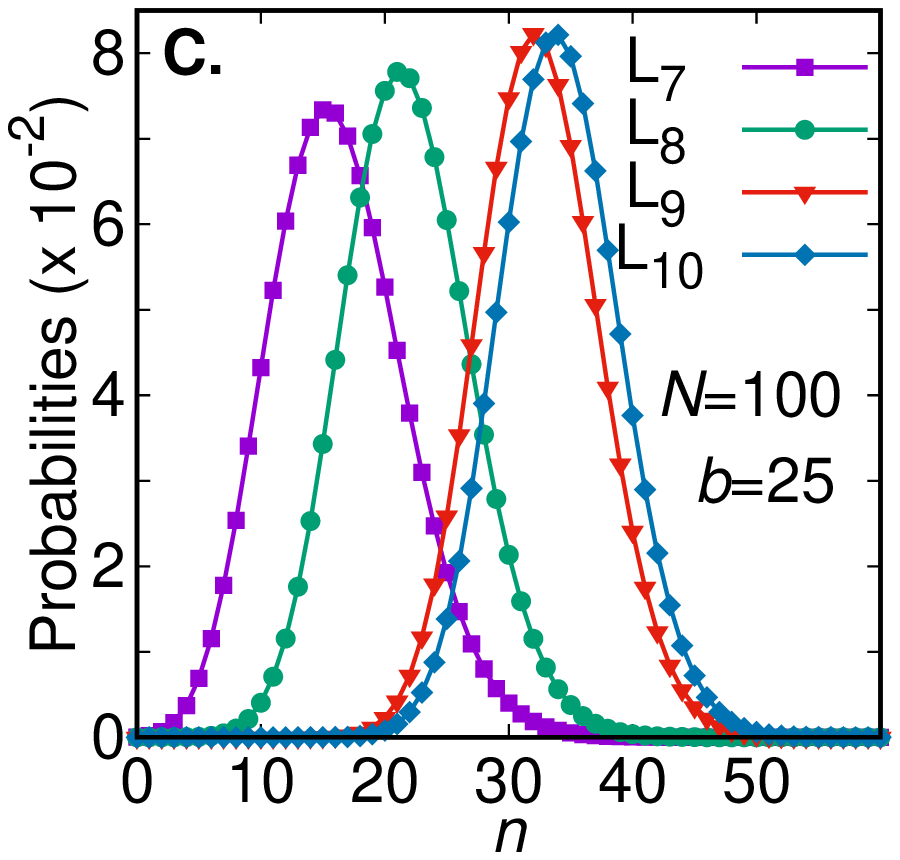}
  \hfill
  \caption{(A) and (B)  show probability distributions obtained
    with $z_0^+$ while $z_0^-$ was used to construct graph
    C. Approximate values of $z_0^\pm$ were obtained from
    Eq. (\ref{eq:z0}). Graph A has $(a,z_0^+)$ in L$_1=(0.07,0.99)$,
    L$_2=(0.099,0.99)$, and L$_3=(0.09,0.99)$. Graph B has $(a,z_0^+)$
    in L$_4=(6,0.71)$, L$_5=(15,0.86)$, and L$_6=(24,0.99)$. Graph C
    has $(a,z_0^-)$ in L$_7=(6,0.35)$, L$_8=(15,0.29)$,
    L$_9=(100,0.13)$, and L$_{10}=(150,0.10)$.}
  \label{fig:probs}
\end{figure}

A third type of splitting arises from the following.  The parameter
$a$ is the eigenvalue of the Cartan operator and gives the OFF to ON
transition rate (see Eq. (\ref{eq:car1})). The action of the ladder
operators on the probability generating function $\phi_{b,a}$ changes
the value of $a$ by one (Eq. (\ref{eq:lad1})) and connects probability
distributions in which $b$ values are the same and $a$ values differ
by an integer.  The action of the raising operator changes
$a=\frac{f}{\rho}$ to $a'=a+1 \rightarrow \frac{f'}{\rho'} = 1+\frac{f}{\rho}$,
and the action of $E_-$ is constructed by analogy. Let us assume that
the action of $E_+$ only changes $f$, hence $\rho'=\rho$ and
$f'=f+\rho$.  $b$ remains unchanged under the action of $E_+$, hence
the remaining constants $N$ and $z_0$ % (or, alternatively, $h$) must
change. For a fixed value of $z_0$ one has $N\rightarrow N -
\frac{1}{1-z_0}$. For a fixed value of $N$ we consider that ${z_0^\pm}
\rightarrow z_0^\pm + \Delta z_0^\pm$ with $\Delta z_0^\pm = \pm
(1+\frac{1}{2N})\sqrt{1-\frac{4bN}{(N+a+1)^{2}}} \mp
\sqrt{1-\frac{4bN}{(N+a)^{2}}}$. The increment of $ z_0^+ $ (or $
z_0^- $) corresponds to an decrease (or increase) of the value of $h$
that implies an increase of the mean protein number (see FIG
\ref{fig:probs}B).

The dynamics of the gene expression process may have two distinct
characteristics, depending on the value of $b$. For $b \gg 1$ the
dominant decay rate to equilibrium is $\rho$ and the changes of the
value of $h$ are not sufficient to cause changes in the time for the
system to approach equilibrium,  $b(\rho+h)$. This regime
coincides with a unimodal probability distribution and the action of
the raising operator on the generating functions causes the mode of
its probability distribution to be displaced to the right. For the
case of $b(\rho+h)\ll \rho$ (or $b\ll z_0 $) we have that $b(\rho+h)$
is the dominant decay rate and the increase (or decrease) of $h$
corresponds to the system reaching equilibrium earlier (or
later). This regime is characterized by probability distributions that
may become bimodal and the action of the raising operator corresponds
to an increase of the maximum probability of finding $n$ equal to
the higher mode (see FIGs \ref{fig:probs}A and \ref{fig:split}B).

The regime with bimodal distributions has important experimental and
theoretical consequences. In this regime, most of the proteins
synthesized by the gene during the ON state are degraded before it
switches back to the OFF state, and the remaining proteins degrade
before the gene switches ON, giving rise to bimodal distributions of
$n$ which have been experimentally observed
\cite{Suter2011}.  In that case, the assumptions underlying
the Langevin approach fail \cite{Gillespie2000} because the number of
proteins  {$\overline{P_L}$} at the steady state regime
of the Langevin equation is governed by distributions that are
Gaussian around  {$\overline{P}$}. The probability
distributions in this case and the breakdown of the Langevin regime
are shown in FIG \ref{fig:split}B.

We began our treatment by considering the macroscopic system because
the master equation solution applies to cases with any number of
molecules. This point is demonstrated by FIG \ref{fig:split}A, which
shows that increasing the equilibrium binding affinity $K=\frac{h}{f}$
reduces the deterministic equilibrium concentration, as expected. For
fixed $K$, reducing $f$ requires reducing $h$. FIG \ref{fig:split}A
shows that there is a symmetry breaking as the average protein number
in the deterministic model splits from that given by the stochastic
model, and moreover that the average number of proteins in the
stochastic model is a function of $f$ even when $K$ is held constant,
behavior never seen in the deterministic model. Although the
deterministic correspondence principle holds for small numbers of
molecules ($\approx 10$), correspondence is lost at one repressor
molecule per cell, as discussed above.

The kinetic symmetries fully manifest themselves in the macroscopic
case. The invariant of the algebra, $b=\frac{f}{\rho}+\frac{h\,
k}{(\rho+h)^2}$ is the ratio between the switching rate and the
protein removal rate. Since the invariant is quadratic in $h$, there
exist two kinetic regimes for the same value of $b$. The first has
protein removal predominantly because of protein destruction (for
example, when $\rho \gg h$) while protein binding prevails in the
second. These regimes are macroscopically indistinguishable in the
presence of a thermodynamically large number of operator sites. This
fully macroscopic picture in fact never occurs in a biological system,
because the molecular number of operator sites per cell is small. In
the ``semi-macroscopic'' case of many protein molecules and a small
number of operator sites, corresponding to the $z_0^+$ branch in FIG
\ref{fig:z0}A,B, protein removal takes place primarily by first order
decay. This super-Fano regime approaches the solutions of a near
equilibrium thermodynamic system as molecular number increases. In the
$z_0^-$ branch, protein removal takes place primarily by binding, the
operator becomes strongly repressed, and sub-Fano behavior results, a
situation we have discussed in detail elsewhere \citep{Ramos2015}.

A further symmetry breaking manifests itself for certain values of $b$
with respect to the protein number distribution when the number of
operators is small.  For the case of $b<1$ the gene switching is slow
in comparison with the protein removal rate, hence the probability
distributions for the protein number when the gene is ON (or OFF) are
split, and bimodal probability distributions are observed (see FIGs
\ref{fig:probs}A and \ref{fig:split}B). When the operator number is
large, these differences in ON and OFF states would average out in the
reaction mixture and become unobservable. Here the actual biological
regime of small gene number per cell is experimentally significant
because it permits direct observation of stochastic switching between
ON and OFF in living cells \citep{Suter2011}. When the gene switching
is fast in comparison with protein removal rate ($b>1$) the
distributions are unimodal and the existence of the two gene states
cannot be established by the measurement of protein numbers, even with
low gene copy number (see FIG \ref{fig:probs}B).

In conclusion, we have made use of symmetries described by a Lie algebra to
fully characterize the behavior of a self-repressing gene. Because the exact
solutions represent the behavior of the system for any number of reacting
molecules and all values of kinetic constants, we interpret the deviation from
deterministic behavior, the splitting of the two branches of $z_0$, and the
emergence of bimodal protein distributions as different types of symmetry
breaking. The role the symmetries play in this analysis differs from how they
are used in quantum problems, where symmetries involve the quantum state
directly. Deeper insight into the role of symmetries will be helpful not only to
statistical physics, but also to other areas involving stochastic processes,
including biological evolution \cite{Hornos1993,Bashford2004}.

\acknowledgments{AFR was supported by CAPES (88881.062174/2014-01). JR was supported by NIH R01
  OD010936. We thank UnJin Lee and David H. Sharp for helpful comments, and
  J. E. M. Hornos for stimulating discussion.}

\bibliographystyle{unsrt}

\end{document}